\documentstyle[12pt,aps]{revtex}

\title{\bf Four Quantum Conservation Laws for Black Hole 
            Stationary Equilibrium Radiation Processes} 
\author{S. Q. Wu\thanks{E-mail: sqwu@iopp.ccnu.edu.cn} 
    and X. Cai\thanks{E-mail: xcai@wuhan.cngb.com}  \\
   \footnotesize \sl Institute of Particle Physics, Hua-Zhong 
                     Normal University, Wuhan, 430079, China } 

\date{\today}
\begin{document}
\maketitle
\baselineskip 23.5pt
\begin{quote}

The classical first law of thermodynamics for a Kerr-Newman black hole (KNBH) 
is generalized to a law in quantum form on the event horizon. Then four quantum 
conservation laws on the KNBH equilibrium radiation process are derived. The 
Bekenstein-Hawking relation ${\cal{S}}={\cal{A}}/4$ is exactly established. 
It can be inferred that the classical entropy of black hole arises from the
quantum entropy of field quanta or quasi-particles inside the hole. 

Key words: Conservation laws, Hawking radiation, Black hole entropy.

PACS number(s): 04.70.Dy  97.60.Lf
\end{quote}

\vskip 0.4cm

It has been a quarter century since Bekenstein$^{1}$ and Hawking$^{2}$ first 
showed that the entropy of a black hole is one fourth of its surface area. 
Despite considerable effort$^{3}$ on the quantum$^{4}$, dynamic$^{5}$, and 
statistical$^{6}$ origins of black hole thermodynamics, the exact source and 
mechanism of the Bekenstein-Hawking black hole entropy remain unclear$^{7}$.

By using the brick wall model, G 't Hooft$^{8}$ identified the black hole
entropy with the entropy of a thermal gas of quantum field excitations outside 
the event horizon, whereas Frolov and Novikov$^{4}$ argued that the black hole entropy 
can be obtained by identifying the dynamical degrees of freedom of a black 
hole with the states of all fields which are located inside the black hole. 
A black hole acts as classical thermodynamic object, but its true microscopic 
structure is unknown$^9$. 

Here we first derive the thermal spectrum and microscopic entropy of a massive 
complex scalar field on a Kerr-Newman black hole (KNBH) background. From this 
quantum entropy, we propose a quantum first law of black hole thermodynamics. 
Then we consider a system in which a KNBH is in equilibrium with this scalar 
field. Using a thermodynamic method, we obtain four conservation laws on black 
hole thermal radiation equilibrium process for energy, charge, angular 
momentum, and entropy, respectively. The total quantities of these quantum 
numbers of the whole system are conserved in the stationary thermal equilibrium 
radiation process. By identifying the complex scalar field with quasi-particles 
excited by the hole, we propose that the classical entropy of a black hole 
originates microscopically from the entropy of quanta which constitute the hole.

The general stationary axisymmetric solution to the Einstein equation is a 
rotating charged black hole (KNBH) described by three parameters: mass $M$, 
charge $Q$, and specific angular momentum $a=J/M$. So we deal with a sourceless 
charged massive scalar field with mass $\mu$ and charge $q$ on this background 
in the non-extreme case $(0<\varepsilon=\sqrt{M^2-a^2-Q^2}\leq M)$; we use 
Planck units $G=\hbar=c=k_B=1$).

In Boyer-Lindquist coordinates, a complex scalar wave function $\Psi$ has 
a solution in variables separable form$^{10}$: 
\begin{equation}
 \Psi(t,r,\theta,\varphi)=R(r)S(\theta)e^{i(m\varphi-\omega t)} 
\end{equation}
Here the angular wave function $S(\theta)$ is an ordinary spheroidal function
with spin weight $s=0$ which satisfies the Legendre wave equation$^{11}$:
\begin{equation}
 \frac{1}{\sin\theta}\partial_{\theta}[\sin\theta\partial_{\theta}S(\theta)]
 +[\lambda-\frac{m^2}{\sin^2\theta}-(ka)^2\sin^2\theta]S(\theta)=0, 
\end{equation}

\noindent
while the radial wave function $R(r)$ is a modified generalized spheroidal 
wave function with an imaginary spin weight which satisfies the following 
"modified" generalized spin-weighted spheroidal wave equation of imaginary 
number order$^{12,13}$: 
\begin{eqnarray}\nonumber
 \partial_r[(r-r_+)(r-r_-)\partial_rR(r)]+[k^2(r-r_+)(r-r_-)\\
 +2(A\omega-M{\mu}^2)(r-M)+\frac{[A(r-M)+\varepsilon B]^2}{(r-r_+)(r-r_-)}+
\\ \nonumber (2{\omega}^2-{\mu}^2)(2M^2-Q^2)-2qQM\omega-\lambda]R(r)=0,
\end{eqnarray}
where $\lambda$ is a separation constant, and $r_{\pm}= M\pm\varepsilon, k^2
={\omega}^2-{\mu}^2, A=2M\omega-qQ, \varepsilon B=\omega(2M^2-Q^2)-qQM-ma.$ 

When considering the thermal radiation of a KNBH, we need the asymptotic 
solutions of the radial function $R(r)$ at its event horizon $r=r_+$. In 
fact, the radial equation has two solutions whose indices at its regular
singularity $r=r_+$ are $\pm iW$, where $W$ is given blew. These two 
asymptotic solutions are
\begin{equation}
 R(r) \sim (r-r_+)^{\pm iW}, \hskip 0.5cm {\rm when} 
 \hskip 0.5cm r \rightarrow r_+
\end{equation}

According to the analytical continued method suggested by Damour and 
Ruffini$^{14}$, these two solutions differ by a extra factor $e^{2\pi W}$. 
It is easy to obtain a thermal radiation spectrum$^{15-16}$ on the event 
horizon $r=r_+$:
\begin{equation}
  \langle N \rangle =\frac{1}{e^{4\pi W}-1}, \hskip 0.5cm
 W=\frac{\omega-m\Omega-q\Phi}{2\kappa}.
\end{equation}
where the surface gravity is $\kappa=(r_+-M)/{\cal{A}}$, the angular velocity 
is $\Omega=a/{\cal{A}}$, the electrical potential is $\Phi=Qr_+/{\cal{A}}$,
and the reduced horizon area is ${\cal{A}}=r_+^2+a^2$. 

Eq.(5) demonstrates that a KNBH has an exact thermal property characterized 
by a temperature $T=\kappa/(2\pi)$ as common blackbody radiation does. The 
radiation modes of a complex scalar field are characterized by a frequency 
$\omega$, a charge $q$ and an azimuthal quantum number $m$. This scalar field 
is rotating with an azimuthal angular velocity $\Omega$ and has a chemical 
potential $\Phi$. The emitted scalar quanta obey Bose-Einstein statistics. 
Here, we adopt Bellido's$^{15}$ proposition that quantum number $W$ is the 
quantum entropy of scalar fields on the KNBH background. The quantum entropy 
satisfies Bekenstein's$^1$ first law of black hole thermodynamics: 
\begin{equation}
 \omega=2\kappa W+m\Omega+q\Phi. 
\end{equation}
and we call this law as the quantum first law of black hole thermodynamics 
in integral form.

When a KNBH is in thermal equilibrium with a complex scalar field at 
temperature $T=\kappa/(2\pi)$, we could regard the hole's surface gravity, 
angular velocity, and electrical potential as external parameters that remain 
fixed or at most undergo a minute change which can be neglected under our 
present consideration. This assumption gives the following conditions of 
thermodynamic stable equilibrium of the system on the event horizon:
$$ \kappa_{r_++0}=\kappa_{r_+-0}, \Omega_{r_++0}=\Omega_{r_+-0},
 \Phi_{r_++0}=\Phi _{r_+-0}, $$

In this thermodynamic equilibrium system, the hole still emits and absorbs
quanta although its parameters remain fixed. However, this stationary process 
is a detailed balance process$^{17}$, that is, the number of quanta emitted 
by the hole is equal to that absorbed by it. So the hole could preserve its
parameters unchanged. Relative to the fixed parameters of the hole, due to
vacuum polarization, the scalar field has some minute fluctuations which can
be given by differentiating the energy relation of Eq.(6). Thus, we have the 
quantum first law of quantum thermodynamics in differential form:
\begin{equation}
 \Delta\omega=2\kappa\Delta W+\Omega\Delta m+\Phi\Delta q.
\end{equation}

Combining Eq.(7) with  the following classical first law of black hole 
thermodynamics in differential form$^{17}$ 
\begin{equation}
 \Delta M=\frac{\kappa}{2}\Delta{\cal{A}}+\Omega\Delta J+\Phi\Delta Q,
\end{equation}
one can deduce four quantum conservation laws for energy, angular momentum, 
charge, and entropy, respectively:
\begin{eqnarray}
  \Delta M &=& n\Delta\omega  ({\rm Energy})  \\
  \Delta J &=& n\Delta m  ({\rm Angular \hskip 2pt Momentum}) \\
  \Delta Q &=& n\Delta q   ({\rm Charge}) \\
  \Delta{\cal{A}}/4 &=& n\Delta W  ({\rm Entropy}) 
\end{eqnarray}
Here $n$ is a multiplier needed to be determined further.
 
Eqs.(9-12) indicate that a KNBH has discrete increments of energy, angular 
momentum, charge, and entropy. That is, when a black hole emits particles, 
its energy, charge, angular momentum, and entropy are carried away by these 
quanta, and vice versa. These microscopic laws are only the reformulated 
detailed balance principle on stationary equilibrium process of black hole 
radiation. In a stationary thermal equilibrium radiation process, it is 
reasonable physically to conceive that what the hole gains meets with the ends 
of that the radiation loses. Thus, the total quantities of energy, charge, 
angular momentum, and entropy of the whole system remain conserved in this 
thermodynamic process.

However, this thermal equilibrium is in general unstable due to the existence 
of statistical fluctuations$^{18}$. A minute perturbation will result in the 
hole completely evaporating to scalar field quanta or radiation quanta being
absorbed fully by the hole. In the former case, the energy, charge, angular
momentum, and entropy in the whole system will convert to those of the scalar 
field, in the latter case to those of the hole. But these physical quantities 
should be equal in these two extreme cases. From this, we could infer that a 
black hole consists of some elementary quasi-excitations, although at present
we do not know what they really. In this paper, we relate them to the scalar 
field quanta. When considering all modes of field excitations, the above 
conservation relations (9-12) must include summation with respect to all 
possible modes of field quanta. 

Further, combining Eq.(6) with integral Smarr formulae$^{19}$
\begin{equation}
 M=\kappa{\cal{A}}+2J\Omega+Q\Phi,
\end{equation}
we can obtain a special quantum state $nm=J, n\omega=M/2, nq=Q/2, nW={\cal{
A}}/4$. As quantum numbers $m, \omega, q, W$ are discrete numbers, not only 
the parameters $J, M, Q, {\cal{A}}$, but also $\Delta J, \Delta M, \Delta Q, 
\Delta{\cal{A}}$ must take discrete values. This means that a quantum 
KNBH can be thought of microscopically as consisting of all possible 
quasi-particles inside the hole, which is identified by us with all 
possible modes of bosonic field quanta having energy $2\omega$, charge 
$2q$, angular momentum $m$, and entropy $4W$ as elementary units.

In fact, Eq.(12) is a generalized second thermodynamic law in quantum form.
By integrating this equation, we obtain the quantum black hole entropy:
\begin{equation}
   nW=\frac{1}{4}{\cal{A}}+C.
\end{equation} 

As the Bekenstein-Hawking classical black hole entropy$^{1,2}$ is $S=A/4
=\pi{\cal{A}}$, the quantum entropy $nW$ is equal to the reduced entropy 
$nW={\cal{S}}=S/(4\pi)$, so we have the Bekenstein-Hawking relations ( 
choose constant $C=0$):
\begin{equation}      
  {\cal{S}}=nW={\cal{A}}/4.
\end{equation}

Eq.(15) shows that the Bekenstein-Hawking black hole entropy is equal to the
quantum entropy of a complex scalar field. On these grounds, one can conjecture 
that the classical entropy of black holes originates statistically from the
quantum entropy of quantized fields.

In summary, we have considered the thermodynamics of a system consisting of a 
complex scalar field in thermal equilibrium with a Kerr-Newman black hole.  
Using the thermodynamic equilibrium condition on the event horizon, we derive 
four quantum conservation laws for black hole equilibrium radiation processes. 
The total energy, total charge, total angular momentum, and total entropy of 
the whole system are conserved in this process. By identifying the interior 
structure of a KNBH with a collection of quasi-particles, we infer that the 
classical entropy of a black hole originates microscopically from the quantum 
entropy of quanta inside the hole. However, this is still an open question 
that needs to be clarified. 

\vskip 4pt
This work was supported in part by the NSFC (No. 19875019) and Hubei-NSF in 
China.

\begin{center}
{\bf \Large References}
\end{center}

\begin{enumerate}
 \item J. D. Bekenstein, Phys. Rev. {\bf D 7}, 2333 (1973); {\bf D 9}, 3292 
 (1974).
 \item S. W. Hawking, Commun. Math. Phys. {\bf 43}, 199 (1975).
 \item J. Bekenstein, {\sl Do we understand black hole entropy ?}, 
 gr-qc/9409015.
 \item L. Bombelli, R. Koul, J. Lee, and R. Sorkin, Phys. Rev. {\bf D 34}, 373
 (1986).
 \item V. Frolov and I. Novikov, Phys. Rev. {\bf D 48}, 4545 (1993), 
 gr-qc/9309001;
  V. Frolov, Phys. Rev. Lett. {\bf 74}, 3319 (1995), gr-qc/9406037.
 \item W. H. Zurek and K. S. Thorne, Phys. Rev. Lett. {\bf 54}, 2171 (1985).
 \item M. H. Lee and J. Kim, hep-th/9604130; Phys. Lett. {\bf A 212 }, 323
 (1996), hep-th/9602129;
  J. Ho, W. T. Kim, Y. J. Park, and H. Shin, Class. Quant. Grav. {\bf 14}, 
2617 (1997), gr-qc/9704032. 
 \item G. 't Hooft, Nucl. Phys. {\bf B 256}, 727 (1985);
  L. Susskind and J. Uglum, Phys. Rev. {\bf D 50}, 2700 (1994), hep-th/9401070. 
 \item M. Maggiore, Nucl. Phys. {\bf B 429}, 205 (1994), gr-qc/9401027. 
 \item B. Carter, Phys. Rev. {\bf 174}, 1559 (1968).
 \item P. Moon and D. E. Spencer, {\sl Field Theory Handbook}, 2nd version,
 (Springer-Verlag, New York, 1971);
  P. M. Morse, H. Feshbach, {\sl Methods of Theoretical Physics}, (McGraw-Hill,
 New York, 1953);
  {\sl Handbook of Mathematical Functions}, edited by M. Abramowitz and I. A. 
 Stegun, 9th version, (Dover, New York, 1972).
 \item S. Q. Wu, X. Cai, J. Math. Phys. {\bf 40}, 4538 (1999), gr-qc/9904037.
 \item E. W. Leaver, J. Math. Phys. {\bf 27}, 1238 (1986).
 \item T. Damour and R. Ruffini, Phys. Rev. {\bf D 14}, 332 (1976).
 \item J. G. Bellido, hep-th/9302127.
 \item G. W. Gibbons and M. J. Perry, Proc. R. Soc. Lond. A {\bf 358}, 467 
(1978).
 \item D. Bekenstein and A. Meisels, Phys. Rev. {\bf D 15}, 2775 (1976). 
 \item L. Liu, {\sl General Relativity} (Advanced Education Press, Bejing, 
1987).
 \item L. Smarr, Phys. Rev. Lett. {\bf 30}, 71, 521(E) (1973).
\end{enumerate}

\end{document}